# The Arrow of Time is Alive and Well but Forbidden Under the Received View in Physics

Ruth E. Kastner

*University of Maryland, College Park,
United States*

**Abstract.** This essay offers a meta-level analysis in the sociology and history of physics in the context of the so-called "Arrow of Time Problem" or "Two Times Problem," which asserts that the empirically observed directionality of time is in conflict with physical theory. I argue that there is actually no necessary conflict between physics and the arrow of time, and that the observed directionality of time is perfectly consistent with physics unconstrained by certain optional metaphysical, epistemological and methodological beliefs and practices characterizing the conventional or Received View.

## 1. Introduction

Scientific research always takes place against a backdrop of particular ground-level assumptions. Such assumptions can be roughly categorized into those that are primarily (i) ontological, (ii) epistemological, and (iii) methodological. There is interaction among these categories; for example, particular ontological assumptions will dictate what is to be viewed as appropriate methodological approaches. (An illustration: it is now considered an established ontological fact that the Earth is round, so a theoretical model involving a flat Earth is viewed as methodologically inappropriate and not considered.) The set of such assumptions, along with the class of theories that is seen as comporting appropriately with them, can be described as the Received View of science at any particular epoch.

The current Received View of physics comprises (at least) four belief-assumptions.[1] I suggest categorizations for these foundational beliefs, but these are not necessarily definitive. They are:

1. The measurement problem of quantum theory is either in-principle unsolvable, or is not really a problem (i.e., not a disqualifying anomaly of standard quantum theory itself). Thus, models that claim to solve the measurement problem should be treated with a high degree of suspicion. (epistemological/methodological).

---

[1] There are at least 2 others, which are not directly relevant to the current topic but are addressed in a separate work (Kastner and Schlatter, 2023). These are:

5. Thermodynamic entropy is equivalent to epistemic uncertainty. (epistemological/methodological)

6. For thermodynamic purposes, measurement is copying. (methodological)

2. Field propagation is always unitary and causal (mechanistic metaphysics; so-called "action at a distance" is forbidden). (Standard quantum field theory picture.) (ontological/methodological)
3. Physical theory, being time-reversal invariant, conflicts with the human experience of an "arrow of time" ("Two Times Problem"). (epistemological/methodological)
4. All physical systems exist in a spacetime container or against a background spacetime; acceptable theories must presuppose this background as fundamental. (A form of *actualism*) (ontological/methodological)

## 2. Critique of the Core Beliefs of the Received View

In this section, I subject the above Beliefs to a critical evaluation. I do not pretend that the evaluation is in any way exhaustive, since the topic under study is obviously of significant scope. However, the following considerations are the ones most relevant to the current issue, namely the alleged discrepancy between physical theory and the "human experience of time". In what follows, I abbreviate each of the Beliefs as B1, B2, etc.

**2.1 Belief 1:** The measurement problem of quantum theory is either in-principle unsolvable or not to be considered a disqualifying anomaly for the standard theory.

The Measurement Problem (MP) can be defined as the inability of the standard formulation of quantum theory to specify what physical interaction counts as a "measurement" warranting attribution of an outcome-eigenstate to a measured system S, such that S can be accurately represented by a proper mixed state (as opposed to an improper mixed state resulting from tracing over other degrees of freedom). The MP is a direct consequence of B2, which assumes that the only *physically quantifiable* field propagation is unitary (and thus deterministic). While we will address B2 in more detail below, it plays a crucial role in the persistence of B1 and therefore the two unavoidably overlap and must be discussed together.

The most famous illustration of the MP is the Schrödinger Cat Paradox, which demonstrates the consequences of presumed-unbroken unitary evolution. The Cat Paradox was presented as a *reductio ad absurdum* of the theory by Schrodinger, although its original function—to demonstrate an anomaly of the standard theory—has largely been obscured. The assumed persistence of unitary evolution (B2) is a bedrock premise of the standard formulation of quantum theory (QT), whether or not QT is taken as involving a "collapse postulate" (CP). This is because a CP is not supported by any quantitative physical account under B2. Therefore, when the CP is invoked, it is an *ad hoc* instrumental device for consistency with the empirical data of determinate outcomes, which instantiate the Born Rule.[2] However, the Born Rule itself does not arise from the presumed ongoing physical unitarity, and thus is also an *ad hoc* device in the standard theory.[3]

This situation leads to well-known inconsistencies (e.g., Frauchiger and Renner 2019) constituting problematic anomalies for the standard theory (Kastner 2020a, 2021, 2023). While B1 generally isn't stated explicitly, it is implied by a number of well-established behaviors among researchers. Among these are:

a) extant attempts to portray standard quantum theory's inconsistency anomalies arising from the measurement problem as counterintuitive truisms about the world (e.g., Proietti *et al*. 2019, Bong et al. 2020, Brukner 2015, Baumann and Wolf 2018). This arguably constitutes anomaly neglect, cf. Kastner (2023), Gradowski (2022: iv, 154).

b) ongoing neglect of an alternative formulation of quantum theory that offers a specific solution to the MP, along with a physical derivation of the Born Rule, but which denies aspects of the Received View.

Behavior (a) evades the theoretical inconsistencies arising from the MP by re-interpreting them not as shortcomings of the theory but as putatively valid predictions of incommensurable differences among different observers. Instances of (a) are critically discussed, for example, in Kastner (2021) and (2023a). These works point out that it is in fact untenable to use the inconsistencies as arguments for a conclusion that QT predicts incommensurable differences among different observers, because contrary to the usual assumption in discussions of the Frauchiger-Renner (FR) inconsistency, such observers could in fact communicate their experimental results, making these and their associated probabilities empirically comparable. In other words, the inconsistent results cannot be considered private and incommensurable in support of claims such as "different observers see irreconcilable facts"; instead, the standard theory itself (i.e., the theory under B2) generates empirically consequential inconsistencies. In view of this situation, the most charitable conclusion for the standard theory is that one of the observers must find that he is mistaken about his state assignment even though he has applied the theory in the standard way (i.e., upon obtaining an outcome, applying the relevant eigenstate to the system). The less charitable conclusion is that the theory is simply incoherent and unable to yield a consistent prediction at all. Even under the charitable conclusion, the theory is rendered a flawed instrument that may require updating of incorrect state assignments.

Regarding behavior (b), which evinces a methodological choice on the part of the mainstream physics community: a solution to the measurement problem is generally eschewed if it does not subscribe to the Received View, and as noted above, there cannot be an internally consistent solution to the measurement problem (at least not one fully consistent with the Born Rule of standard quantum theory)[4] under B2 of Received View. However, there does exist a specific solution which rejects B2, namely the Relativistic Transactional Interpretation (RTI; e.g., Kastner 2022). RTI has been available for over a decade (with a precursor form available since the 1980s),[5] but it has thus far not been admitted into mainstream discussion as an eligible approach. Gradowski (2022: 6, 154) discusses this phenomenon of epistemic insensitivity of the mainstream community to competitor theories.[6] Of course, the MP is not the focus of the current work, which concerns the arrow of time; nevertheless, the latter is not really separable from the topic of the MP, insofar as the transactional (RTI) solution to the MP yields not only the Born Rule but also a physically real arrow of time, consistent with our empirical experience of change (this point is reviewed in Section 3). Thus, entrenchment of the Received View leads to multiple obstacles and problems that are indeed readily resolved upon relinquishment of inappropriate metaphysical beliefs and their attending methodological impediments.

---

[2] This *ad hoc* character is reflected in the so-called "Shifty Split" in which there is no principled physical account of which systems are to be taken as described by quantum theory under B2. On one side of the "split" the theory is simply suspended without theoretical justification. The resulting lacuna is often covered by recourse to "consciousness," but this is unsatisfactory since it leads to the Wigner's Friend-type inconsistencies.

[3] The Born Rule was famously arrived at as an educated guess by Max Born (1926).

Before turning to the remainder of the Received View beliefs, we should acknowledge here that of course there are many who would claim that their preferred interpretation of the standard formulation of quantum theory (which supervenes on B2) does solve the MP. For example, it is often claimed that the Bohmian hidden variable theory (BHV) or Everettian ("Many Worlds") approaches solve the MP.[7] However, these approaches remain subject to the empirically fatal inconsistencies arising from the Wigner's Friend scenarios such as Frauchiger-Renner (2019) (again, see Kastner 2020a, 2023a, for why these inconsistencies are empirically consequential). Since these inconsistencies arise directly from the standard theory's inability to define the conditions for the occurrence of an outcome *accurately meriting an eigenstate description* — which is the MP— these approaches fail to solve the MP despite their provision of narratives that seem to support the occurrence of an outcome.

**2.2 Belief 2:** Field propagation is always unitary; "action at a distance" is forbidden.

We now turn to further specifics of Belief 2, which stems from the metaphysical commitment to a mechanistic notion of causality inherited from classical physics. This is what we might call the "bucket brigade" (BB) picture, in which it is assumed that any influence among systems must always be communicated locally via a mediating field at sub-light or light speed in a future-directed process. This belief is implemented in current practice by way of quantum field theory (QFT), in which it is postulated that a quantum field operator be associated with every spacetime point, acting as an element of a "bucket brigade" that carries conserved quantities "through spacetime" from a source to a sink in a time-asymmetric way via the Feynman "causal" propagator exclusively. The mediating field is viewed as a necessary physical structure for the propagation of any physical influence. No other sort of propagation is permitted under this belief. In particular, so-called "action at a distance" is viewed as particularly offensive and scientifically unacceptable. However, this belief turns out to be inconsistent with B3. (We will see why when we deal with the specifics of B3 in §2.3.)

---

[4] E.g., so-called "spontaneous collapse theories" such as the Ghirardi-Rimini-Weber (1986) theory, which add an *ad hoc* term to the Schrodinger evolution, deviate from the Born Rule.

[5] E.g., Cramer (1986).

[6] Gradowski's observation (2022: 154) is especially relevant in this regard: "The inability or unwillingness of the mainstream to entertain promising alternative theories can lead to both theory-ladenness and anomaly neglect or even blindness, as when continental fixists called the jigsaw-fit of the continents an illusion."

[7] An argument by Norton shows that decision-theoretic derivations of the Born Rule, such as those appealed to under the Everettian approach, are circular since they presuppose the universal necessity of probabilities at the outset. (Norton 2021, Chapter 10). See also Kastner (2016) for a circularity problem in the Everettian approach.

Underpinning B2 is the traditional classical adherence to a Democritan mechanistic metaphysics. Examples of the commitment of prominent physicists to this metaphysical picture are the following statements by Christian Huygens and William Thomson:

> "In the true philosophy we can see the causes of all natural effects in terms of mechanical motions. And, in my opinion, we must admit this, or else give up all hope of even understanding anything in physics." (Huygens, *Traité de Lumière*, 1690)

> "It seems to me that the test of 'Do we or do we not understand a particular topic in physics?' is 'Can we make a mechanical model of it?'" (Thompson, *Notes of Lectures on Molecular Dynamics*, 1885).

However, it should be noted that Heisenberg's breakthrough in QT began only when he gave up on precisely such mechanical models and instead allowed the empirical data to direct him to the correct form of the theory, which decidedly conflicts with the local realism implied by classical mechanistic metaphysics (cf. Kastner 2022, Chapter 2). This historical fact should provide at least some modest support for questioning the above mindset, however esteemed were the classical physicists who subscribed to it. Yet it appears clear from current practice of insisting on a local, mechanical model of quantum fields (and more generally any physical process taken as legitimately explicated) that B2 is still in full force. Einstein's abhorrence of "spooky action at a distance" remains in play to this day, as we see in ongoing attempts to "save locality" in QT (cf. Henson 2017, Kastner 2017b, and Kastner 2011 for a critical discussion of such attempts).

**2.3 Belief 3:** Physical theory, being time-reversal invariant, conflicts with the human experience of an "arrow of time."

Turning now to B3: time-reversal invariance is the property in which a theoretical quantity does not change upon reversal of the sign of the time index. Of course, this belief directly gives rise to the "Arrow of Time" problem. The view that physical law is time-reversal invariant is often stated as if it were established fact, for example in Buonamano and Rovelli (2023) and by Carroll (2010). Carroll says:

"The weird thing about the arrow of time is that it's not to be found in the underlying laws of physics. It's not there. So it's a feature of the universe we see, but not a feature of the laws of the individual particles. So the arrow of time is built on top of whatever local laws of physics apply."[8]

However, the above statement, however categorically stated and consensus-endorsed, turns out to be false: the theoretical requirements for the propagation of conserved currents (such as energy) as well as the standard approach to QFT, both of which apply at the level of individual particles, *are time-asymmetric*. We explicate this point further in §2.3.B. For now, we note that the propagators of relativistic quantum theory are all explicitly temporally directed and decidedly not time-reversal invariant. Examples are the retarded, advanced, and Feynman propagators. Thus QFT reflects an arrow of time, as it must in order to propagate real conserved currents (see §2.3.B).

Belief 3 thus appears mistakenly to take time-reversible Newton mechanics (i.e., neglecting any dissipative forces) as representative of all physical law applicable to individual systems. This situation naturally gives rise to consternation about how to reconcile physical theory with the "human experience of a flow of time"; i.e., to the so-called "Two Times Problem" (TTP). Since the Received View precludes a resolution of this discrepancy in terms of putative physical law, proposed resolutions attribute it to illusory or otherwise non-veridical features of human experience (i.e., aspects of psychology or neurology) or to *ad hoc* perspectival assumptions (e.g., we are "moving through" a putative block world).[9] The latter begs questions such as "Why and how are life forms 'moving through' the purported block?", and "Why and how do some systems count as 'moving through' (e.g., humans) and not other systems (e.g., rocks)?"

*2.3.A Belief 3 involves a double standard concerning "empirical" data*

This situation raises a concern regarding what sorts of phenomena count as "veridical," and at a more basic level, what counts as empirical data serving as the standard for such a criterion. Physics is first and foremost an empirical science, in that its theories must be supported by empirical data and at the very least, must be consistent with –as in, not contradict– empirical data. In this regard, it should be noted that considering the human dynamical experience of time as illusory in order to support beliefs such as B3 effectively contradicts the empirical data, which is, e.g.: "the system was at $x_1$ at time $t_1$ and at $x_2$ at time $t_1$", where those events were experienced as temporally separate (i.e., not simultaneously). Let us call those sorts of empirical observations *change-data*, where the term is defined by different states instantiated by a system in an experienced temporal sequence. That is what was seen and what can be corroborated among different observers; even though they may use different coordinate systems, no observer can dispute the change *empirically*. Thus, taking change-data as non-veridical constitutes a redefinition of what counts as "empirical data" in order to uphold beliefs such as B3, which is not actually based on empirical data but on certain theoretical or metaphysical models. Yet, alas, the theoretical models (such as wave equations) propping up B3 cannot actually support real energy propagation (this point is discussed in §2.3.B below).

---

[8] https://www.wired.com/2010/02/what-is-time/
[9] See Kastner (2023b), footnote 2 for why purported empirical demonstrations of the truth of the perspectival account are unsound since they amount to affirming the consequent.

Furthermore, this tacit redefinition of what counts as "empirical" constitutes a double standard. As is well known, in the late 19th century it was believed that classical physics was nearing a final triumph in successfully accounting for all the empirical data available. However, the two "small clouds" on the horizon were empirical data that classical physics failed to predict: (1) the blackbody frequency spectrum and (2) discrete atomic spectral lines. The physics community correctly concluded that classical physics was inadequate, since it was inconsistent with the empirical data in these cases. (As noted earlier, Heisenberg only made progress in developing QT when he abandoned classical mechanical model-making –i.e., he rejected B2 as exemplified by the views of Huygens and Thompson). What we have today in the form of the so-called "Two Times Problem" is change-data that similarly refutes certain core beliefs. However, instead of discarding those inconsistent core beliefs, as was done for example in Heisenberg's successful development of QT, the prevailing approach is to disqualify change-data from being the empirical data that it in fact is, and instead to declare it illusory or otherwise ineligible to qualify as scientific data. That this constitutes a double standard is clear from the fact that at no time did physicists conclude that the phenomena of atomic spectral lines were illusory features of human subjective experience, and attempt to look for psychological or neuroscientific explanations for these because classical physics –then the Received View– was inconsistent with the data. Since ultimately, all empirical data is "subjectively experienced," the double standard is to disqualify corroborable change-data while accepting other types of corroborable data, such as the blackbody spectrum and atomic spectral lines.

We can entertain a possible objection to the assertion that change-data is empirical data, as follows. The objector could attempt to claim that certain kinds of "empirical data" have been shown to be illusory, as for example in the case of the appearance of the Sun going around the Earth, and that change-data can therefore be illusory in the same manner. However, this would be a false equivalence. The idea that the Sun goes around the Earth is not the empirical data itself, but rather what seems to be an obvious *inference* from the data. The data itself is nothing more than: at a time $t_1$ the Sun was at location $x_1$ (in some local coordinate system) and at time $t_2$ the Sun was at location $x_2$. That data remains valid regardless of our theory about which body is going around which, and indeed the corrected theory –that it is the Earth that is going around the Sun– *is consistent with the same data* while better accommodating additional data (observed planetary motions). Change-data is precisely the same sort of basic empirical data as above. It is neither a theory nor a metaphysical assumption.[10] Thus, this objection fails.

---

[10] Of course, data can in principle be theory-laden. However, the only theory-ladenness in this sort of data is the invoking of some coordinate system, and we can recognize that the coordinate system is not absolute without discounting the data.

In view of the above considerations, the "Two Times Problem" can be understood as a pseudo-problem that arises only from critical defects in the Received View, and that in fact physical theory is perfectly consonant with our empirical experience of change and becoming, i.e., an Arrow of Time. This is a case that has been made in the literature before, in various forms (e.g., Reichenbach (1953, 1956)[11], Sorkin (2003), Kastner (2012, 2017a, 2022, 2023a), Schlatter and Kastner (2023), Schlatter (2021); but the relevant points have been disregarded or overlooked (again, apparently due to entrenchment of the Received View, cf. Gradowski 2022: iv, 154). They are reviewed again in Section 3, but the interested Reader is invited to peruse the relevant references for further details.

*2.3.B. Belief 3 is contradicted by current theory*

While Carroll's remark at least acknowledges that change-data are empirical data, his assertion that temporally oriented behavior is "not a feature of the laws of the individual particles" (i.e., B3) is falsified by the following facts:

(a) B3 does not actually describe current practice in the theoretical description of individual parties; and
(b) B3 actually precludes any unitarily propagating field as assumed in B2.

Specifically: the Feynman propagator $D_F$ of standard QFT is not time-reversal invariant, as it describes positive energy propagating in the positive temporal direction and negative energy propagating in the negative time direction. It explicitly breaks time symmetry. Thus, standard QFT is, by construction, *not time-reversal invariant*. The only time-reversal invariant propagator is the time-symmetric propagator $\bar{D}$, which is the most general form of field propagation (both positive and negative energies in both temporal directions). However, $\bar{D}$ is the principle part of a complex function (e.g. Davies, 1971).[12] This means that it cannot propagate any real (on-shell) energy, and thus cannot support unitary evolution understood as the propagation of any on-shell field excitation: $\bar{D}$ corresponds only to virtual (off-shell) fields. *The only way that propagation of real energy can occur is if time-symmetry is broken.* I italicize the previous point since it is generally not recognized, although the breaking of time symmetry is nevertheless imposed "by hand" in the service of upholding other foundational beliefs such as B2. In standard QFT this consists of imposing a "causal" propagator, such as the Feynman propagator, as an *ad hoc* choice driven by B2.

Thus, standard practice is to break time symmetry "by hand," at the level of individual particles; and thus B3 –the belief that the applicable physical laws are time-symmetric at the level of individual particles– is unambiguously false. The foregoing demonstrates that the Received View has self-contradictory aspects, such that subscribing to it necessitates overlooking these inconsistencies. In particular, one cannot claim that the laws of physics are time-reversal invariant (B3) while also insisting on a mechanistic model that propagates energy in a preferred temporal direction (B2). Of course, as noted above, we must have temporal symmetry breaking in order to have any propagation of real, on-shell energy, and that is why we can indeed have a real Arrow of Time consistent with physical theory. But this does not require a mechanistic metaphysics, and it can indeed allow for a form of time-symmetry to play a supporting, if not universal, dynamical role. We deal with this issue further in Section 3.

---

[11] In fact, Reichenbach argued that physical causation, in the sense of coming into being, was relevant to the Lorentz invariance of time order for timelike or null-separated events (Reichenbach 1956: 25).

[12] Specifically, the Green's function f(x) for a quantum field has the general form $f(x) = 1/x \mp i\pi\delta(x)$, and the time-symmetric propagator $\bar{D}$ is the real part of this function, which does not include on-shell energies. The latter appear in the delta function, i.e., in the imaginary part.

**2.4 Belief 4:** All physical systems exist in a spacetime container or against a background spacetime.

We turn now to Belief (4), the assumption that to be physically real, an entity must be an element of the spacetime manifold. This assumption is reflected in the reference literature, for example in this excerpt from the Stanford Encyclopedia entry on Quantum Field Theory:

> "But is a systematic association of certain mathematical terms with all points in space-time really enough to establish a field theory in a proper physical sense? *Is it not essential for a physical field theory that some kind of real physical* properties *are allocated to space-time points*? This requirement seems not fulfilled in QFT, however. Teller (1995: ch. 5) argues that the expression *quantum field* is only justified on a "perverse reading" of the notion of a field, since no definite physical values whatsoever are assigned to space-time points. Instead, quantum field operators represent the whole spectrum of possible values so that they rather have the status of observables (Teller: "determinables") or general solutions. *Only a specific* configuration*, i.e., an ascription of definite values to the field observables at all points in space, can count as a proper physical field*." (Meinard, 2023)

The latter excerpt features two key assertions of Belief 4, which I have italicized above: (i) the tacit ontological/theoretical requirement of a spacetime background or container, along with (ii) the declaration that anything physically real be an element of that manifold. Specifically: in (ii), the author declares that the requirement for a "real physical" entity is that it have definite values against this assumed spacetime background.[13] The above excerpt serves to illustrate the prevailing established view that the term "physically real" is equivalent to "is an element of the spacetime manifold," where the latter is understood as the delimiter of physical existence. This can be understood as a form of *actualism.* However, my use of the term is not an endorsement of the conventional definition of that term as the opposite of *possibilism,* where the latter is defined in terms of possible world semantics. In my usage, an *actual* is simply an element of the spacetime manifold, i.e., an invariant *event.* Such entities are to be distinguished from other real physical entities and processes –such as quantum systems and their interactions– that are not elements of the spacetime manifold. The latter can be understood as *physical possibilities* or a form of *res potentia* (Kastner, Kauffman, Epperson 2018) that are the ontological referents of QT. Thus, QT can be understood as describing a substratum for spacetime. While this is of course a startling idea given our phenomenal impression that we live in a spacetime container, there are compelling theoretical reasons to accept it; we briefly touch on these below. (And it is worth mentioning in this regard that it was equally startling to consider the idea that the Earth was actually going around the Sun.)

---

[13] Technical point: "Having the status of observables" in the excerpt glosses over the fact that field creation and annihilation operators are not actually observables, since they are not operators on Hilbert or Fock space, which is the domain of the field excitation states constituting quanta. The latter are characterized by number operators, the legitimate observables. Thus the Received View (which includes standard quantum field theory) has no theoretical place for the quantum field operators; they are not quanta in Hilbert space nor are they elements of spacetime. Yet standard QFT defines its basic theoretical entity in terms of such operators. This inconsistency is remedied in the heterodox direct-action picture of fields (Kastner, 2015).

While in recent years some researchers have been entertaining the possibility that spacetime as a continuous entity or delimiter is not fundamental (as discussed for example in Becker, 2022), the theories under consideration still presuppose a form of spacetime actualism. In particular, theories such as Loop Quantum Gravity tacitly assume that QT involves a spacetime background. However, the latter is a wholly optional metaphysical premise that is actually in tension with QT, as follows: relativistic quantum theory demotes spacetime indices to mere parameters, and there is no relativistic position operator nor time operator at any level of quantum theory. These are sound reasons to question the conventional assumption that QT implies or requires the ontological existence of a spacetime background. In contrast, conserved currents such as energy and momentum retain their observable status at all levels. This means that quantum systems retain the corresponding eigenstate descriptions independently of any spacetime properties, and in that sense the conserved currents are more physically fundamental than the spacetime construct.[14] In any case, at the relativistic level, the conserved currents are unambiguously "preferred observables" since the spacetime quantities are not observables at all.[15] In this picture, the conserved currents physically give rise to emergent spacetime symmetries, not the other way around as traditionally assumed in discussions of Noether's theorem (cf. Schlatter and Kastner 2023).

The consequences of B4, which basically consists in denying ontological existence to anything that does not constitute an element of spacetime (another consequence of Democritan metaphysics), are far-reaching. In particular, they severely restrict the debate around discussions of temporality, and have done so for centuries, even before the advent of relativity unified metrical space and time.[16] Relativity theory, which simply provides a covariant map of the metrical features of the spacetime manifold, is now uncritically taken as delimiting the entire territory of physical existence in keeping with the longstanding entrenchment of B4. This belief system leads to conclusions of an incompatibility of temporal passage or "becoming" with physical theory, as in the following:

> "If on the other hand the world *is* fundamentally temporal in the way that Sellars insists it must be, then (at least as far a special relativity as a representation of that world is concerned), Sellars' famous scientific realism is compromised because what he deems for metaphysical reasons to be an 'essential feature' of a temporal picture of the world does not appear in a fundamental theory of spacetime" (Savitt, 2021)

---

[14] Of course, the uncertainty principle still applies, but this means that one can have an exact momentum eigenstate along with completely indeterminate position, while the inverse –an exact position eigenstate and completely indeterminate momentum– is not really physically possible, since there are no real position eigenstates. Traditional references to position "values" are limited to a non-relativistic approximation and cannot legitimately be used to support ontological claims regarding spacetime.

[15] This shows how neglect of the relativistic level in discussions of the MP impedes its solution, since one alleged feature of the MP is basis arbitrariness. But such arbitrariness is restricted to the nonrelativistic theory, which is only an approximate limit. It also reveals a weakness in the usual decoherence arguments which, in seeking to explain the appearance of locality, help themselves to the position basis when there is actually no position observable. In contrast, the appearance of locality is fully accounted for in RTI through the localizing effect of emission and absorption of conserved currents (Kastner 2020b).

In the above, "the world" (presumably denoting the domain of all that exists) is tacitly taken as equivalent to (or at least delimited by) spacetime. Thus the excerpt illustrates the usual ground-floor assumption of B4 –that there can be nothing ontologically real outside the spacetime manifold. However, one can reject B4 and instead consider what we call "spacetime" (really, a set of invariant events) to be a subset of the world's ontology, namely the realm of actuals, while allowing in addition the existence of ontological potentialities described by QT. Then, as Reichenbach suggested, *"The flow of time is a real becoming in which potentiality is transformed into actuality,"* (Reichenbach 1953). (This transformation of quantum potentiality into spacetime actuality is quantitatively described in Schlatter and Kastner (2023) and yields a novel theory of quantum gravity in which QT and relativity are harmoniously reconciled.)

Thus, unburdened by B4, one can indeed be fully scientifically realist and still assert that the world has temporal features. In this regard, it is instructive to recall Ernan McMullin's observation that the realist is not restricted to conventional categories:

> [I]maginability must not be made the test for ontology. The realist claim is that the scientist is discovering the structures of the world; it is not required in addition that these structures be imaginable in the categories of the macroworld. (McMullin 1984: 14)

The well-known ostensible tension of QT with relativity is a manifestation of B4, which effectively implies that quantum theory must not be referring to something real if that to which it refers seems to "violate" relativity theory. Once B4 is laid down, relativity theory and QT are not in conflict, since neither is presumed to exclusively describe "the world." And indeed their domains of applicability, as well as the relationship between the two theories, can be quantitatively and consistently delineated. This has already been demonstrated in the literature, e.g., Kastner (2022), Chapter 8; and, as noted above, Schlatter and Kastner (2023).

---

[16] Prior to relativity, the delimiting container or background was the Democritan "void" in an actualist metaphysics.

Moreover, even if one does not wish to consider any sort of "becoming" picture, B4 cannot legitimately be maintained even under standard general relativity (GR). This is because GR implies that matter sources of the spacetime manifold –which is the metric field of GR– are outside the field. Any curvature resulting from the existence of a mass M does not vanish anywhere in the vicinity of M, and therefore M cannot be part of the field –i.e., cannot be part of spacetime. As alluded to earlier, material quantum systems –which are the sources of the spacetime metric field– can be understood as elements of a *quantum substratum* (Kastner 2022, Chapter 8). This again is fully consistent with the fact that field sources are not elements of the fields they generate –they are always outside the field. For example, electric charge, as a source of the electromagnetic field, is not part of the electromagnetic field. We have no difficulty conceiving of the latter because we don't think of the electromagnetic field as the delimiter of all that physically exists. Spacetime, as a metric field generated by matter sources, is no more a delimiter of the physically real than is any other field. Thus the "masses in the void" Democritan metaphysical presumption of B4 now essentially has the status of a folk belief, since it is refuted by GR, even though that has generally not been noticed. This sort of blindness to the implications of theories, whether they be anomalies or unexpected ontological/structural aspects, is a drawback of belief entrenchment of the kind discussed in Gradowski (2022).

## 3. The Arrow of Time can be found in physical law unburdened by inappropriate metaphysical beliefs

We have seen in the foregoing that the standard assertion that the arrow of time is "not to be found in the underlying laws of physics" is simply untenable, both in view of standard theoretical practice and in the requirement for the propagation of real (on-shell) conserved currents. The latter cannot occur without temporal symmetry breaking, i.e., irreversibility. This instructs us that there must be such a breaking of time symmetry in Nature in keeping with the empirical facts observed, and in particular with the apparent predictive success of the "causal" Feynman propagator in standard QFT, which is required for the propagation of real energy, momentum, and angular momentum.

In the context of non-relativistic QT, the claim that the basic laws of physics are time-reversible comes from the tradition of eschewing real reduction, i.e., irreversible "collapse" in QT (i.e., from adherence to B2). However, if there is such collapse in Nature, that provides the missing link between an underlying time-symmetric physical theory and the empirically established fact of an arrow of time. In the formulation mentioned previously, the Relativistic Transactional Interpretation (RTI), one has exactly this. While quantitative details of RTI can be found in the References cited in the foregoing, we briefly review the basics here as a qualitative overview. RTI is the relativistic elaboration of the Transactional Interpretation (TI) originally proposed by Cramer (1986). TI is based on the Wheeler-Feynman "absorber theory" or "direct action theory" (Wheeler and Feynman 1945, 1949). Let us denote the Wheeler Feynman Classical Absorber Theory as WFCAT to emphasize that *it is a classical theory only*, and as such does not fully describe the quantum level. (The quantum form of the direct-action theory is described in quantitative detail in Kastner (2022), Chapter 5). WFCAT involves a basic time-symmetric interaction between charged currents, which at the relativistic level can be described by the time-symmetric propagator $\bar{D}$. An additional process is that of "absorber response," in which absorbers generate their own $\bar{D}$ in response to the emitter. The sum of these fields is a causal field in the form of the retarded propagator in the classical version and the original TI. The WFCAT theory treated absorber response as universal, but it is crucial to recognize that this is not the case at the quantum level.

The quantum relativistic development, RTI, improves upon this basis by providing a quantitative account of the advent of "absorber response", which is taken as primitive in the original 1986 version. Absorber response is a non-unitary process which at the relativistic level transforms the time-symmetric propagator $\bar{D}$ into the Feynman propagator $D_F$. The interaction referred to as "absorber response" in the original TI is actually a relativistic interaction in which both emitter and absorber(s) (electromagnetic field sources) mutually participate. This process is fully quantified in RTI and essentially corresponds to decay rates in the standard theory (Kastner 2022, Chapter 5; Kastner and Cramer, 2018). Under RTI, the time-symmetric propagator $\bar{D}$ is identified as the Coulomb force involved in scattering (non-radiative) processes, while the imaginary part of the Feynman propagator $D_F$ arising from the mutual non-unitary interaction is identified as a real photon, i.e., a radiative process.

Thus, under RTI we have a time-symmetric theoretical foundation (in the form of the time-symmetric propagator $\bar{D}$) along with a specific mechanism for temporal symmetry breaking, which in addition to establishing a temporal arrow via the Feynman propagator $D_F$, yields objective reduction. At the relativistic level, the objective reduction corresponds to the transformation of $\bar{D}$ into $D_F$, so that that latter need not be imposed in an *ad hoc* manner as in standard QFT. The latter point was first observed by Davies (1970, 1971, 1972) and is further elaborated in Kastner (2022, Chapter 5). It is also noted by Breuer and Petruccioni (2000), who point out that the imaginary part of the temporally directed Feynman propagator $D_F$ can be understood as a decohering factor arising from absorber response in the direct-action theory fields.[17] Since absorber response is a non-unitary process, it violates B2. However, the action corresponding to the Feynman propagator is already non-unitary (Breuer and Petruccione 2000; Kastner 2022, Chapter 5). Thus, non-unitarity supporting a criterion for "measurement" is ironically already present in the standard theory despite the mainstream adherence to B1 and B2. This inconsistency illustrates, yet again, the unsupportable status of the Beliefs of the Received View despite their entrenchment.

An additional feature of RTI is that it accepts that quantum entities –systems described by Hilbert or Fock space vectors– are not elements of spacetime. This is a requirement for a realist approach to QT, since if quantum states are taken as referring to something physically real, that to which they refer cannot be a component of 3+1 spacetime, since Hilbert or Fock space elements are multidimensional and complex. Thus, RTI implicitly violates B4 (but so, also, does standard GR, as noted previously). An important dividend of RTI is a quantitatively well-defined account of "measurement", i.e., a solution to the MP (e.g., Kastner 2017c, 2020b, 2023a). But since the latter violates B1, it is been seen not as a dividend but instead as a methodological violation; thus the formulation has not been admitted into mainstream discussion.[18]

---

[17] See Kastner (2020b) for a quantitative discussion of decoherence under RTI corresponding to objective reduction.

In any event, the RTI formulation straightforwardly yields an arrow of time and predicts precisely the same phenomena (e.g., propagation of conserved currents, outcomes in accordance with the Born Rule) as the standard theory. It improves on the standard theory by deriving the Born Rule from primary physical principles rather than postulating it, as a dividend of its solution to the MP. The existence of this formulation (as well as the fact that standard QFT possesses an arrow of time, however *ad hoc* the latter might be) falsifies the conventional claim that physical law is at odds with the directionality of time and the associated change-data.

## 4. Conclusion

This work has argued that the so-called "Arrow of Time Problem" or "Two Times Problem" (TTP) is essentially a myth. It is an article of faith that is contradicted by actual theoretical practice, in which the applicable physical theory does possess an Arrow of Time and must do so in order to account for the propagation of real energy and other conserved currents. Belief in the TTP is upheld only through entrenched adherence to a set of primarily metaphysical beliefs of a predominant Received View of physics that themselves are contradicted by the empirical facts, by current theory, and by inconsistencies among the beliefs themselves.

Moreover, the inadequacies in current standard theory –such as the MP, the *ad hoc* nature of the temporal symmetry breaking in standard quantum field theory, and logical inconsistencies pointed to by Haag's theorem (such as the lack of a basis for interacting quantum fields)– are all remedied by an extant formulation, the Relativistic Transactional Interpretation (RTI). RTI is empirically equivalent to the standard approach (unlike the *ad hoc* "spontaneous collapse" theories currently admitted to mainstream discussion) and provides a more quantitative and precise foundation for the empirically observed Arrow of Time. However, due consideration of this remedy by the mainstream community, which has not occurred to date, apparently requires a willingness to recognize and to question the firmly held Beliefs of the Received View.

---

[18] The reader might object that other "spontaneous collapse" interpretations, such as the Ghirardi-Rimini-Weber theory (1986) have become part of mainstream discussion. However, these are frankly *ad hoc* approaches that change the Schrodinger equation and deviate from the Born Rule. In this respect they conform to B1, in that B1 holds that "real" quantum theory cannot have non-unitary collapse. RTI challenges that belief while GRW-type approaches do not. When objective reduction approaches are considered at all in mainstream discussion (e.g., Nurgalieva and del Rio (2018), GRW-type approaches are taken as representative of all such approaches, which effectively serves both to misrepresent and suppress RTI.